\documentstyle[preprint,epsfig,aps]{revtex}

\epsfverbosetrue

\newcommand{\br}{\mbox{$\bf r$}}
\newcommand{\bR}{\mbox{$\bf R$}}

\newcommand{\bG}{\mbox{$\bf G$}}
\newcommand{\bk}{\mbox{$\bf k$}}
\newcommand{\ci}{\mbox{$\rm i$}}
\newcommand{\calH}{\mbox{$\cal H$}}

\newcommand{\beq}{\begin{equation}}
\newcommand{\enq}{\end{equation}}

\newcommand{\hbttm}{\mbox{$\frac{\hbar^2}{2m}$}}

\draft

\begin{document}
\title{Spectral Shifts of Semiconductor Clusters}
\author{Antonietta Tomasulo and Mushti V. Ramakrishna}
\address{The Department of Chemistry, New York University,
New York, New York 10003-6621.}
\date{Submitted to Surface Review and Letters, \today}
\maketitle

\begin{abstract}

The shifts of the electronic absorption spectra of GaAs and GaP
semiconductor clusters are calculated using accurate pseudopotentials.
In the absence of experimental data at present, these calculations
provide estimates of expected spectral shifts in these clusters.  In
addition, these calculations show that Coulomb interaction between the
electron and hole dominates over the confinement energy in small
clusters, with the result that the electronic absorption spectra of
small clusters exhibit redshift instead of blueshift as the cluster
size is decreased.

\end{abstract}
\pacs{PACS numbers: 71.35.+z, 36.20.Kd, 36.40.+d}

Semiconductor clusters exhibit electronic and optical properties quite
different from those of molecules and solids
\cite{Pool:90,Corcoran:90}.  For this reason, there have been several
spectroscopic studies on these systems, designed to elucidate the
evolution of their optical properties with cluster size
\cite{Rosetti:83,Brus:86,Wang:87,Bawendi:89,Bawendi:90,Shiang:90,Yoffe:93,Hu:90}.
The ultimate benefit gained from such studies is the detailed and
microscopic insights into the size dependence of the cluster
properties.  Such a knowledge may enable us to engineer materials with
applications in microelectronics and computer industry.

In support of these experimental efforts, we recently developed a band
structure model for calculating the spectral shifts of semiconductor
clusters as  a function of cluster size \cite{RK:91,RK:92}.  The
spectral shift refers to the shift of the leading edge of the
electronic absorption spectrum of a semiconductor cluster as the
cluster size is varied.  Our calculated spectral shifts were in
excellent agreement with experimental values for CdS clusters \cite{RK:91}.  We
also
carried out similar calculations on GaAs and GaP clusters \cite{RK:91}.
However,
all those calculations employed local empirical pseudopotentials.
Since non-local pseudopotentials are more accurate than the local
pseudopotentials, we now performed calculations on GaAs and GaP clusters
using non-local pseudopotentials.  The results of these calculations
are reported here.

Our electronic structure calculations utilize the empirical
pseudopotential method (EPM) that has been previously used for the
investigations of the optical properties of bulk semiconductor
materials and clusters \cite{RK:91,RK:92,Chelikowsky:76,Cohen:89}.
This method consists of solving the Schr\"{o}dinger
equation using an empirically determined pseudopotential for the valence
electron,
\beq
\calH = -\hbttm\nabla^2 + V_p,
\enq
\beq
V_p(\br, E) = V_L(\br) + \sum_{l = 0}^{\infty} \Pi_l^{\dagger} A_l(E)
f_l(\br)\Pi_l,
\enq
where the first term is the purely local part, the second term gives
the non-local $(V_{NL})$ part, and $\Pi_l$ is the projection operator
for angular momentum states $l$.  The function
$A_l(E)$ is the energy dependent well depth and $f_l(\br)$ is conveniently
taken to be the square well
\beq
f_l(\br) = \left\{ \begin{array}{ll}
                   1 \hspace{4.0 em} &  \br < R_l \\
                   0 \hspace{4.0 em} &  \br \geq R_l.
                   \end{array}
           \right.
\label{flr}
\enq
The local part of the pseudopotential is given by
\beq
V_L(\br)
=  \sum_{\bG} \left[ V_S(G)S_S(\bG) + \ci V_A(G)S_A(\bG) \right]
\exp(\ci\bG\cdot\br),
\enq
where $V_S$ and $V_A$ are the symmetric and anti-symmetric form
factors, respectively, determined by fitting them to the experimental
optical data.  Similarly, $S_S$ and $S_A$ are the symmetric and
anti-symmetric structure factors, respectively, determined from the
crystal structure \cite{Cohen:89}.  The local EPM employs only
$V_L(\br)$ in its Hamiltonian.

To apply EPM to the calculation of the semiconductor clusters we assume
that these clusters have the crystal structure of the bulk
semiconductor.  This assumption is justified because we are considering
relatively large clusters containing tens to hundreds of atoms.
Indeed, these clusters are large enough that
they are characterized by their size rather than
by the number of atoms.
Furthermore, the X-ray and TEM experiments on CdS, CdSe, and Si
clusters have shown that the bulk lattice structure is preserved even
when the cluster radii are as small as \bR~ = 7 \AA
\cite{Brus:86,Wang:87,Bawendi:89,Bawendi:90,Shiang:90,Yoffe:93,Littau:93}.
The probable reason that even these small clusters assume bulk crystal
structure is that the dangling bonds on the surfaces of these  clusters
are passivated by the organic ligands attached to them.

In bulk semiconductors the allowed wave vectors \bk~ of the
electronic states are continuous.  On
the other hand, only discrete $\bk$-states are allowed in clusters.  If
we model the cluster as a rectangular box with dimensions $L_1, L_2$,~
and $L_3$, then a reasonable
approximation is to use
the bulk pseudopotential $V_p(\br)$ inside the
cluster and terminate this potential at the surface by
an infinite potential.  The wave vectors of the lowest
allowed states are then given by the
quantization condition $\sin(k_xL_1)\sin(k_yL_2)\sin(k_zL_3) = 0$,
whose solution is
\beq \bk  =
\pi\left(\frac{n_x}{L_1},\frac{n_y}{L_2},\frac{n_z}{L_3}\right),
\label{krect}
\enq
where $n_x$, $n_y,$ and $n_z$ are the quantum numbers
of a particle in a box with infinite potentials at the boundaries.
For
the low energy excitations under consideration, the assumption of
infinite potentials at the boundaries is a good approximation.
The energy levels at these allowed \bk-states constitute the band structure
of a rectangular box.

Similarly, if we model the cluster as a spherical object
of radius $R$, the energy levels of the valence electrons will be
quantized because of the spherical boundary.  The wave vectors of the
lowest allowed states are given by $j_0(k_nR) = 0$, whose solution is
$\bk_n = n\pi/R$ \cite{Flugge}.  Since $\bk_n$ is along the radial
direction, we project it onto each of the cartesian axes with equal
magnitude to obtain cartesian components of \bk.  This procedure yields
\beq
\bk  = \frac{\pi}{\sqrt{3}R}\left(n_x,n_y,n_z\right).
\label{ksphere}
\enq
The energy levels at these \bk-states constitute the $l = 0$ band
structure of the clusters.  Modification of this formalism to finite
depth potentials at the boundaries is also possible.  However, the
choice of the well depth remains arbitrary and hence we do not employ
such potentials here.

We calculated the band structures of GaAs and GaP clusters by
diagonalizing $137\times 137$ Hamiltonian matrices constructed as
outlined above, with \bk~ given by eq. (\ref{ksphere}).  We employed
the square well non-local potential Eq. \ref{flr} for GaAs, instead of
the Gaussian well employed in Ref. \cite{Chelikowsky:76}.  In addition,
the spin-orbit term in GaAs pseudopotential \cite{Chelikowsky:76} is
omitted and the remaining parameters adjusted to obtain good
bulk band structure.  These two modifications simplified the GaAs
pseudopotential Hamiltonian without affecting its accuracy.  The band
structures of GaAs crystal obtained using local and non-local EPM are
displayed in Fig. 1 along with some experimental data.  The non-local
EPM band structure is seen to agree better with experimental data than
the local EPM band structure.

{}From the band structures of the clusters thus calculated
we determined the energy differences between the highest occupied
(HOMO) and the lowest unoccupied (LUMO) orbitals.  These are
the band gaps
$(E_g)$ of the clusters.   We corrected these band gaps
to take into account the  electron-hole Coulomb attraction
and correlation energies, to obtain\cite{Brus:86,Kayanuma:88}
\beq
E_x = E_g - \frac{1.786 e^2}{\epsilon R} - 0.248 E_{Ry},
\label{Ex}
\enq
where $E_x$ is the spectral shift (or exciton energy) of the cluster
of radius R, $\epsilon$ is the bulk dielectric constant, and $E_{Ry}$
is the effective Rydberg energy of the exciton.

The virtue of EPM is that it reproduces the bulk band structures and
band gaps to better than 0.03 eV accuracy \cite{RK:91,RK:92}.
Specifically, our predicted indirect band gap of 0.43 eV for a 10 \AA~
radius silicon cluster was found to be in good agreement with recent
experimental value of 0.5 eV obtained by Louis Brus and co-workers
\cite{Littau:93}.  Likewise, EPM yielded excellent results for CdS
clusters in comparison with experiment \cite{RK:91,RK:92}.  Other
methods are less accurate, especially for the calculations of the band
gaps.  For example, the calculations based on the density functional
theory within the local density approximation (LDA) typically
underestimates the band gaps by about 30-50\% \cite{Cohen:89}.
Furthermore, the LDA calculations are computationally far more
expensive compared to EPM.  Consequently, EPM is suitable for the
investigation of the electronic structures of semiconductor clusters.
Full details of our computational methodology are given in Refs.
\cite{RK:91,RK:92}.

The spectral shifts of GaAs and GaP clusters thus calculated are
displayed in Figs. 2 and 3, respectively.  Since GaP is an indirect gap
semiconductor, we present spectral shifts of both the direct and
indirect transitions.  The results obtained from the use of both local
and non-local EPM are shown in these figures.
{}From these figures it is clear that over a large
range of cluster sizes the local EPM is able to reproduce spectral
shifts as accurately as non-local EPM.  However, at small cluster sizes
the non-local correction on the spectral shifts is significant.
Furthermore, at large cluster sizes the absorption spectrum shifts to
higher energies with decreasing cluster size.  This blueshift is
expected due to confinement of the electron-hole pair in the cluster.
However, at small cluster sizes the absorption spectrum of GaAs
clusters shifts to lower energies with decreasing cluster size; a trend
opposite to that observed for large clusters.  In the case of GaP,
which is an indirect gap semiconductor, the lowest energy transition
exhibits blueshift at both large and small cluster sizes.  But this
transition is not observable because it is forbidden.  The origin of
the absorption spectrum, corresponding to the observable direct
transition, shifts to lower energies with decreasing cluster size at
small cluster sizes.  Both local and non-local pseudopotentials exhibit
the same qualitative behavior.

We can explain the observed trends in the following way.  At large
cluster sizes the electron and hole are both confined in a spherical
well.  This quantum confinement increases the band gap with decreasing
cluster size and it is the dominant effect in this size regime.  In
these large clusters, the negatively charged electron and the
positively charged hole are spatially separated and hence the coulomb
attraction between them is negligible.  However, in small clusters the
coulomb attraction energy between the electron-hole pair cannot be
neglected.  While the band gap still increases with decreasing cluster
size, in small clusters this increase is sufficiently overcome by the
coulomb energy that the spectra shift to lower energies.  Consequently,
in this small cluster size regime the absorption spectra of clusters
may exhibit redshift instead of the blueshift.

At present reliable experimental data on these systems are not present,
partly because of considerable experimental difficulties that arise in
trying to synthesize III-V semiconductor clusters with narrow size
distribution.  However, our calculations provide reasonable estimates
of the expected spectral shifts of these clusters.

In summary, our band structure calculations of GaAs and GaP
semiconductor clusters have shown that local empirical pseudopotentials
are reasonably accurate compared to the non-local empirical
pseudopotentials for the calculations of the spectral shifts of these
clusters.  The non-local corrections on the spectral shifts are most
important in small cluster sizes.  In addition, our calculations have
shown that, while quantum confinement energy is the dominant factor
affecting spectral shifts in large clusters, the Coulomb interaction
between the electron and hole has significant effect in small
clusters.  The attractive Coulomb interaction is sufficiently strong in
small clusters that it overcomes the confinement energy of the
electron-hole pair and gives rise to redshift, instead of the
blueshift, of the electronic absorption spectrum.

This is the sixth paper in this series on optical properties of
semiconductor clusters.  This research is supported by the New York
University and the Donors of The Petroleum Research Fund (ACS-PRF Grant
\# 26488-G), administered by the American Chemical Society.

\begin{figure}
\caption{The band structure of zinc-blende GaAs crystal obtained using
non-local (solid line) and local (dashed line) pseudopotentials.  The
experimental points are also plotted for comparison.  The non-local
pseudopotential calculations are seen to agree better with the
experimental data than the local pseudopotential calculations.}
\end{figure}

\begin{figure}
\caption{The calculated spectral shifts or exciton energies of
zinc-blende GaAs clusters for low energy excitations.  The open circles
are obtained using local pseudopotentials  and filled circles are
obtained using non-local pseudopotentials.} \end{figure}

\begin{figure}
\caption{The calculated spectral shifts or exciton energies of
zinc-blende GaP clusters for low energy a) direct and b) indirect
excitations.  The open circles are obtained using local
pseudopotentials and filled circles are obtained using non-local
pseudopotentials.} \end{figure}

\newpage

\end{document}